\documentclass[singlespacing]{elsart}

\usepackage{graphicx}

\usepackage{amssymb}
\journal{}
\begin{document}

\begin{frontmatter}

\title{Model for crankshaft motion of protein backbone in nonspecific binding site of serine proteases}

\author{A.E. Sitnitsky},
\ead{sitnitsky@mail.knc.ru}

\address{Kazan Institute of Biochemistry and Biophysics, P.O.B. 30, Kazan
420111, Russia. e-mail: sitnitsky@mail.knc.ru }

\begin{abstract}
The consequences of recent experimental finding that hydrogen bonds of the anti-parallel $\beta $-sheet
in nonspecific binding site of serine proteases become significantly shorter and stronger synchronously
with the catalytic act are examined. We investigate the effect of the transformation of an ordinary hydrogen bond into a low-barrier one on the crankshaft motion a peptide group in the anti-parallel $\beta $-sheet. For this purpose we make use of a realistic model of the peptide chain with stringent
microscopically derived coupling interaction potential and effective on-site potential. The coupling
interaction characterizing the peptide chain rigidity is found to be surprisingly weak
and repulsive in character. The effective on-site potential is found to be a hard one, i.e., goes more steep than a harmonic one. At transformation of the ordinary hydrogen bond into the low-barrier one the frequency of crankshaft motion of the corresponding peptide group in the anti-parallel $\beta $-sheet is roughly doubled.
\\
\end{abstract}

\end{frontmatter}

\section{Introduction}
The problem of localization and storage of energy by proteins at binding of ligands or protein-protein interactions
is an old one in biophysics (for review see \cite{Zho09}, \cite{Sch09} and refs. therein).
This problem is especially urgent within the context of enzymatic catalysis. How does an enzyme
store and utilize the energy released at substrate binding? The
answer to this question is very poorly understood not only at
quantitative but even at qualitative level. In our opinion recent experimental data on serine
proteases \cite{Fod06} are able to shed light on this problem.

Indeed the high resolution X-ray diffraction studies ($\leq 1.2$\ \AA) of serine protease intermediate structures
revealed that "the strength of the hydrogen bonds between the enzyme and the substrate changed during catalysis. The well-conserved
hydrogen bonds of antiparallel $\beta$-sheet between the enzyme and the substrate become significantly shorter
in the transition from a Michaelis complex analogue ... to an acyl-enzyme intermediate ... presumably synchronously with the
nucleophilic attack on the carbonyl carbon atom of the scissile peptide bond. This is interpreted as an active mechanism that utilizes
the energy released from the stronger hydrogen bonds to overcome the energetic barrier of the nucleophilic
attack by the hydroxyl group of the catalytic serine." \cite{Fod06}. It is worthy to note that these data are in coherence with those of \cite{Bon91} (see also \cite{Fuh06} and refs. therein) that the $\beta-$strand corresponding to residues $214-217$ shifts $\sim 0.8$\ \AA\ toward the inhibitor upon its binding by $\alpha-$lytic proteases. Here one should strictly discriminate the established experimental
fact of the hydrogen bonds shortening and strengthening in the antiparallel $\beta$-sheet from the speculation about the participation of this effect in the mechanism of the catalysis.
The latter is an intriguing hypothesis that still needs to be confirmed. It will not be touched upon in the present paper. Our concern
here is in modeling the consequences
of the hydrogen bonds shortening and strengthening in the antiparallel $\beta$-sheet for its dynamics.

Serine proteases are the most extensively investigated enzymes (for review see \cite{Hed02}, \cite{Pol05} and refs. therein). The greatest wealth
of structural and kinetic data has been obtained for them. That is why  these "working horses" of enzymology are the most suitable objects for
investigating unresolved problems of enzymatic catalysis.
The antiparallel $\beta$-sheet mentioned above forms the nonspecific binding site of serine proteases and is quite similar for all of them.
The differences are in specific binding sites which are of no interest for us here. The antiparallel $\beta$-sheet of the nonspecific binding
site comprises three sequential peptide groups of enzyme backbone which form hydrogen bonds with corresponding peptide groups of the substrate
(see Fig.3 in \cite{Hed02} or a schematic picture Fig.24-9 in \cite{Fin02} or that Fig.11.6 in \cite{Bra99}).
Despite the fact that two peptide chains belong to two different proteins the resulting structure has all features of the antiparallel $\beta$-sheet
and is usually called so without reserve.

Nature of hydrogen bonds in proteins is a matter of controversy from the beginning of protein science and enzymology.
In particular the possibility of the so-called low barrier hydrogen bonds in enzyme active sites is of utmost interest.
The proposal about their role for enzymes was put forth by Gerlt, Clealand and Frey in 1993-1994 (for review see \cite{Cle98}, \cite{Fre04}
and refs. therein). Since then their participation in transition state stabilization is a matter of heated debates \cite{War96}, \cite{Sho04}. Especially that between His and Asp
in the catalytic triad of serine proteases remains to be the subject of numerous studies and non-stopping controversy with arguments pro \cite{Cle92},\cite{Cle921}, \cite{Fre94}, \cite{Fre95}, \cite{Ger97}, \cite{Cas97}, \cite{Cle00}, \cite{Ha00}, \cite{Wes02} and contra
\cite{Fuh06}, \cite{Sho04}, \cite{Tam09}. We do not touch upon
the catalytic triad in the present paper and focus ourselves exclusively on the processes in the nonspecific binding site.
The most salient feature of low barrier hydrogen bonds as compared to ordinary hydrogen bonds is their strength and short distance between hydrogen donor and
acceptor atoms (dDA). We do not involve ourselves in
very intricate classification debates about blurred boundaries between weak, strong, short strong, short ionic, low barrier and very strong hydrogen bonds
(see, e.g., \cite{Kar08}). For convenience we freely divide the hydrogen bonds to strong ones (SHB) and weak ones (WHB). We set that
SHB have energies $\geq 10$ kcal/mol and dDA $\leq 2.5$ \AA\  while WHB have
energies $2\div 5$ kcal/mol and dDA $\geq 2.8$ \AA.
Thus the aim of the present paper is to investigate the transformation of
WHB into SHB in the antiparallel $\beta$-sheet of the nonspecific binding site of serine proteases. To be more  precise we study the effect of
this transformation on the backbone dynamics in the antiparallel $\beta$-sheet.
However it should be stressed that following \cite{Kar08} we believe that there are no well
defined boundaries between different types of hydrogen bonds. In fact we investigate the parametric dependence of the backbone dynamics on the hydrogen bond
energy from the range $2\div 30$ kcal/mol.

Last years are marked by noticeable progress in revealing
subtleties of protein dynamics gained by infra-red (IR)
spectroscopy \cite{Wou02}, \cite{Wou01}, \cite{Ham98},
\cite{Ham00}, \cite{Xie00}, \cite{Xie01}, normal mode analysis
\cite{Yu03}, \cite{Lei02}, \cite{Lei01} and molecular-dynamics
simulations (see the recent reviews \cite{Kar02}, \cite{Kar00}
and refs. therein). There are several analytical models treating  mechanics of peptide
chain backbone \cite{Pey95}, \cite{Zor97}, \cite{Zor99},
\cite{Mer96}. The most interesting type of motion for our context is the so
the so called "crankshaft-like" one.
It means the rocking of the rigid plane of the peptide group due to degrees of freedom of torsional (dihedral)
angles $\varphi_i$ and $\psi_{i-1}$ (see Fig.1). It was
proposed on theoretical grounds from normal-mode analysis \cite{Go76} and
is supported by NMR experiments and molecular
dynamics simulations of protein backbone  \cite{McC77},
 \cite{Lev79}, \cite{Gun82}, \cite{Lev83}, \cite{Del89}, \cite{Cha92}, \cite{Pal92},
\cite{Bru95}, \cite{Fad95}, \cite{Buc99},
\cite{Der99}, \cite{Ulm03}, \cite{Clo04}, \cite{Fit07}.
It is comprehended now as a dominant type of motion for the latter that
"involves only a localized oscillation of the plane of the peptide group. This motion results in a strong anticorrelated motion of the $\Phi$ angle of the $i-$th residue
($\Phi_i$) and the $\Psi$ angle of the residue $i - 1$ ($\Psi_{i-1}$) on the 0.1 ps time scale."
 \cite{Fit07}. Thus the essence of this motion is the so called anticorrelated
motion of the torsional angles $\varphi_i$ and $\psi_{i-1}$
manifested itself in the requirement (see Fig.2)
\[
\phi/2\equiv \Delta \varphi_i = \Delta \psi_{i-1}
\]
In this case the plane of the
peptide group rocks as a whole around some axis $\sigma$ that goes through the
center of masses of the peptide group parallel to the bonds
$C_{\alpha}^{i-1}-C^i$ and $N^i-C_{\alpha}^{i}$ (see Fig.1).
The moment of
inertia of the peptide group relative to the axis $\sigma$
can be easily calculated
to be $I \approx 7.34 \cdot 10^{-39}
g \cdot cm^{2}$ (see Appendix).
Molecular dynamics simulations and NMR experimental data suggest that the character of
the correlation function for the crankshaft motion is decaying
oscillations \cite{Gun82}, \cite{Fit07} but provide characteristics
for them in a very wide range from the subpicosecond
and picosecond time scale \cite{Fit07} to  slower motions on a much larger
time scale from tens of picoseconds to 100 ps and more \cite{Clo04}.
This is presumably a manifestation of the "Russian doll" structure of the conformational
potential for the crankshaft motion when a group of
local minima forms a smooth local minimum and so on. In fact  knowing
the actual values of the frequency for the oscillations and the characteristic time of
their decay for the functionally important crankshaft motion is not
indispensable for the purposes of the present paper. However in our opinion
it is reasonable to assume that the frequency of oscillations of the plane
of the peptide group as a whole for such motion should be at least
 order of magnitude
less than those for high frequency in - plane motions such as,
 e.g., Amide-I ($\sim 1600\ cm^{-1}$). The
choice of the frequency in the
$\omega_0 \sim 100\ cm^{-1}\approx 10^{13}\ s^{-1}$ range enables us to match it
with the amplitudes of rocking of order of several degrees (see (\ref{eq12})
below) in accordance with experimental data \cite{Buc99}, \cite{Ulm03}.

To investigate the effect of the transformation of
WHB into SHB in the antiparallel $\beta$-sheet on the characteristics of the backbone crankshaft motion
we use the model developed earlier \cite{Sit07}, \cite{Sit08}. The latter is a microscopically stringent
model of polypeptide backbone dynamics. It deals with realistic effective on-site potentials and
coupling interaction potentials but at the same time is simple and tractable. The model yields for the
correlation function of the crankshaft motion the exponentially decaying oscillations \cite{Sit08} in
qualitative agreement with the results of molecular dynamics simulations \cite{Fit07}. The model enables us to
elucidate the role of hydrogen bonds in the backbone dynamics.

The paper is organized as follows. In Sec.2 the Hamiltonian of the model is constructed.
In Sec.3 the coupling interaction of two adjacent peptide groups is considered.
In Sec.4 the local on-site potential for the peptide group is considered. In Sec.5 the contribution of hydrogen bonds is separately considered because they are the central point of the present paper.
In Sec.6 the dimensionless characteristics of the effective on-site potential and of the coupling interaction potential are discussed. In Sec.7 the equation of motion is
derived and the correlation function for crankshaft motion is obtained. In Sec.8 the results are discussed and the conclusions are summarized. In Appendix some technical details are presented.

\section{The Hamiltonian of the model}
A schematic picture of a polypeptide chain with all designations
used further is presented in Fig.1. The mutual orientation of two
adjacent peptide groups is characterized by the torsional
(dihedral) angles $\varphi _{i}$ and $\psi _{i}$. The angle
$\varphi _{i}$ characterizes the rotation round the bond
$C_{i}^{\alpha }- N_{i}$ and that $\psi _{i}$ characterizes the
rotation round the bond $C_{i}^{\alpha }- C^{'}_{i}$. Further we
consider the equilibrium dynamics in the antiparallel $\beta$-sheet
for which the equilibrium values of the torsional angles are the
same for all peptide groups (they are listed at the end of the
next Sec.). We consider small deviations $\varphi _{i}(t)$ and
$\psi _{i}(t)$ of the torsional angles from their equilibrium
values ($\varphi _{i}= \varphi _{i}^{0}+\varphi _{i}(t)$ and $\psi
_{i}=\psi _{i}^{0}+\psi _{i}(t)$) with $\left|{\varphi _{i}
(t)}\right| \le 20^{\circ}$ and $\left|{\psi _{i}(t)}\right| \le
20^{\circ}$. At such amplitudes of the deviations the hydrogen
bonds confining the peptide group in the antiparallel $\beta$-sheet are
not broken \cite{Fin02}. Finally we consider a peculiar type of
motion $\psi _{i-1}(t)=\varphi _{i}(t)$ that is namely the crankshaft one. The latter means that
the peptide group rotates as a whole respective some effective
axis $\sigma$ passing through the center of the C$-$N bond
parallel to the bonds $C_{i}^{\alpha }- C^{'}_{i}$ and
$N_{i+1}-C^{\alpha} _{i+1}$ that are assumed to be approximately
parallel ($\angle C_{i}^{\alpha }-C^{'}_{i}- N_{i+1} =113^\circ$
and $\angle C^{'}_{i}- N_{i+1}- C^{\alpha} _{i+1} =123^\circ$
\cite{Can80}). This type of motion is stipulated by the fact that
the peptide group is a planar rigid structure \cite{Fin02},
\cite{Can80}. For the sake of uniformity of designations we
further denote $x _{i}=2\varphi _{i}(t)=2\psi _{i-1}(t)$ (see
Fig.2) and thus
\begin{equation}
\label{eq1} \varphi _{i} = \varphi _{i}^{0} + x _{i}/2 ;\quad
\quad \quad \psi _{i-1} = \psi _{i-1}^{0} + x _{i}/2
\end{equation}
The moment of inertia of the peptide group relative to the axis
$\sigma$ can be easily calculated and is $I \approx 7.34*10^{-39}
g*cm^{2}$ (see Appendix). The Hamiltonian of the polypeptide chain in our model
with nearest neighbor interactions is
\begin{equation}
\label{eq2} H = \sum\limits_{i} {\left\{ \frac{I}{2}\left(
\frac{dx _{i} }{dt} \right)^{2} +U_{loc}(x_{i})+U(x_{i}; x_{i+1})
\right\}}
\end{equation}
Here $U_{loc}(x_{i})$ includes interactions defining the local
potential of the peptide group (namely hydrogen bonds, the
so-called torsional potentials and the van der Waals interaction
of covalently non-bonded atoms of the peptide group with the atoms
of adjacent side chains) defining its separate motion while
$U(x_{i}; x_{i+1})$ includes the coupling interactions (namely the
van der Waals interaction of covalently non-bonded atoms of
adjacent peptide groups i and i+1 and their electrostatic
interaction) intermixing the motions of these groups and leading
to the rigidity of the peptide chain. It should be stressed that
the latter potential also contributes into the separate motion of
the peptide groups. We can define the effective on-site potential
for such motion as
\begin{equation}
\label{eq3} V_{eff}(x_{i})=U_{loc}(x_{i})+U(x_{i}; 0)+U(0;x_{i})
\end{equation}
and the coupling interaction potential as
\begin{equation}
\label{eq4} U(x_{i}; x_{i+1})=U^{vdw}(x_{i};
x_{i+1})+U^{el}(x_{i}; x_{i+1})
\end{equation}
Thus the equation of motion is
\begin{equation}
\label{eq5}
I\frac{d^2x_i}{dt^2}=-\frac{dU_{loc}(x_i)}{dx_i}-\frac{dU(x_{i-1};
x_{i})}{dx_i}-\frac{dU(x_{i}; x_{i+1})}{dx_i}
\end{equation}
In the following two sections we consider in details the functions
$U_{loc}(x_i)$ and $U(x_{i}; x_{i+1})$. At doing it we will
freely pass back and forth between the variables according to the
rule (\ref{eq1}) which can be rewritten as
\begin{equation}
\label{eq6} \bigtriangleup\varphi _{i} =  x _{i}/2 ;\quad \quad
\quad  \bigtriangleup\psi _{i} =  x _{i+1}/2
\end{equation}

\section{Coupling interaction defining the rigidity of a polypeptide chain}
Both interactions in (\ref{eq4}) intermixing the motions of the
adjacent peptide groups and contributing to the coupling
interaction are described by the central potentials
$W^{vdw}_{mn}(R_{mn}(\varphi _{i};\psi _{i}))$ and
$W^{el}_{mn}(R_{mn}(\varphi _{i};\psi _{i}))$ between the atom
$A_{n}$ of the $i$-th peptide group and the atom $A_{m}$ of the
$i+1$-th one with $R_{mn}$ being the distance between the atoms
$A_{n}$ and $A_{m}$. The electrostatic potential is
\begin{equation}
\label{eq7} W_{mn}^{el} \left( {R_{mn}} \right) = \frac{{q_{m}
q_{n}} }{{\varepsilon R_{mn}}}
\end{equation}
\noindent where $q_{m}$ and $q_{n}$ are partial charges on the
atoms $A_{m}$ and $A_{n}$ respectively ($q(N)=-0.28e; q(H)=0.28e;
q(O)=-0.39e; q(C)=0.39e$ where $e=4.8*10^{-10}$ CGS \cite{Can80})
and $\varepsilon $ is the dielectric constant which for protein
interior should be better conceived as some adjustable parameter
($ \approx 2 \div 10$ with 3.5 being commonly accepted value
close to high frequency permeability of peptides) \cite{Can80},
\cite{Fin02}. In some studies $\varepsilon $ is supposed to be
solvent dependent and chosen, e.g., 4 for $CCl_4$, 6$\div$7 for
$CHCl_3$ and 10 for $H_2O$ \cite{Das87}, \cite{Pop97}. For the
van der Waals potential one can choose any of the numerous forms
suggested in the literature, e.g., the Lennard-Jones one (6-12) or
the Buchinghem one (6-exp). In this chapter we use for numerical
estimates the former one
\begin{equation}
\label{eq8} W^{vdw}\left( {R_{mn}}  \right) = - \frac{{A_{mn}}
}{{R_{mn}^{6}} } + \frac{{B_{mn}} }{{R_{mn}^{12}} }
\end{equation}
\noindent with the set of the well known parameters of Scott and
Scheraga (other sets were also verified and found to give similar
results).\\

The distance $R_{mn}$ as a function of the angles $\varphi _{i}$
and $\psi _{i}$ is
\[
 R_{mn} (\varphi _i ,\psi _i ) = \{ p_m^2  + r_n^2  +
 \]
 \[
 2p_m r_n [\sin \theta (\cos \gamma _m \sin \alpha _n \cos \varphi _i -
 \sin \gamma _m \cos \alpha _n \cos \psi _i  ) -
\]
 \begin{equation}
\label{eq9}
 \sin \gamma _m \sin \alpha _n (\cos \psi _i \cos \varphi _i \cos \theta  -
 \sin \psi _i \sin \varphi _i ) - \cos \gamma _m \cos \alpha _n \cos \theta ]\} ^{1/2}
\end{equation}

The interaction potential for both van der Waals and
electrostatic interactions has the form
\begin{equation}
\label{eq10}
 U^{\left\{ {\scriptstyle vdw \hfill \atop
  \scriptstyle el \hfill} \right\}} (x _i ;x _{i + 1} ) =
  \sum\limits_{mn} {W^{\left\{ {\scriptstyle vdw \hfill \atop
  \scriptstyle el \hfill} \right\}} } (R_{mn} (x _i ;x _{i + 1} ))
\end{equation}
\noindent where the summation in n is over atoms in the $i$-th
peptide group and that in m is over atoms in the $i+1$-th one.

In what follows we work with full realistic coupling interaction
potential described above. However for making use of suggestive
analogies with the known literature results obtained on toy
models we calculate the so called coupling constant which is a
key characteristic of the truncated coupling interaction
potential. Expanding the potential we obtain that to the leading
order in the terms $x _{i}\ll 1$ and $x _{i+1}\ll 1$ the rigidity
of the peptide chain is determined by the term of the potential
which intermixes the motion of the adjacent peptide groups
\begin{equation}
\label{eq11} U^{mix}\left( {x _{i} ;x _{i + 1}}  \right) \approx
(-K)x _{i}x _{i + 1}
\end{equation}
where the coupling constant is
\[
-K =-(K^{vdw} + K^{el}) =
\]
\begin{equation}
\label{eq12}\frac{{\partial ^{2}U^{vdw}\left( {x _{i} ;x _{i +
1}}  \right)}}{{\partial x _{i} \partial x _{i + 1}} }_{\left| {x
_{i} = 0;x _{i + 1} = 0} \right.} + \frac{{\partial
^{2}U^{el}\left( {x _{i} ;x _{i + 1}} \right)}}{{\partial x _{i}
\partial x _{i + 1}} }_{\left| {x _{i} = 0;x _{i + 1} = 0} \right.}
\end{equation}

The results of calculations of the contributions into the
coupling constant for the antiparallel $\beta$-sheet are as follows:
 $\varphi _{i}^{0}=-139^{\circ}$;
$\psi _{i}^{0} =135^{\circ}$
\cite{Can80}:$-K^{vdw}\approx-5.0\cdot10^{-16}$ erg;
$-K^{el}\approx2.64\cdot10^{-15}$ erg;
$-K\approx2.14\cdot10^{-15}$ erg at $\varepsilon =3.5$.

\section{Local on-site potential of the peptide group}
The local potential $U_{loc}(x_i)$ is composed as follows
\begin{equation}
\label{eq13}
U_{loc}(x_i)={U}_{sc}(x_i)+{U}_{hb}(x_i)+{U}_{tors}(x_i)
\end{equation}
Here ${U}_{sc}(x_i)$ is the energy of van der Waals interactions
of the atoms of the $i$-th peptide group with those of the side
chains $R_i$ and $R_{i-1}$ and also with the atoms $H^i_{\alpha}$
and $H^{i-1}_{\alpha}$,
${U}_{hb}(x_i)={U}^{(1)}_{hb}(x_i)+{U}^{(2)}_{hb}(x_i)$ is the
energy of two hydrogen bonds of the $i$-th peptide group and
${U}_{tors}(x_i)={U}_{tors}^{\varphi}(\varphi
_{i})+{U}_{tors}^{\psi}(\psi _{i-1})$ is the energy of the
torsional potentials for the rotation of the angles $\varphi
_{i}$ and $\psi _{i-1}$. For the latter we take the usual form
\[
{U}_{tors}^{\varphi}(\varphi
_{i})=E_{\varphi}(1+\cos3\varphi _{i})
\]
\begin{equation}
\label{eq14}{U}_{tors}^{\psi}(\psi _{i})=E_{\psi}(1+\cos3\psi
_{i})
\end{equation}
with $E_{\varphi}\approx$1 kcal/mol and $E_{\psi}\approx$1
kcal/mol. In all further simulations we choose the side chain to
be Ala ($R_i$ is $C^i_{\beta}(H)_3$) and initially find the
value of the angle $\chi_1$ (see Fig.1) to minimize the energy.

\section{Hydrogen bond potential}
In (\ref{eq13}) ${U}^{(j)}_{hb}(x_i)$ ($j$=1,2) are the contributions of two hydrogen bonds
which a peptide group in the anti-parallel $\beta $-sheet forms with its partners.
We start from a model Morse potential usually employed to describe the hydrogen bond energy dependence on its
current length \cite{Das87}, \cite{Pop97}
\begin{equation}
\label{eq15} U_{hb}(x_i)=D\left(1-exp\left[-n(r(x_i)-r_0)\right]\right)^2-D
\end{equation}
Here $n\approx 3$ \AA$^{-1}$ and $r_0\approx 1.8$ \AA\  are adjustable parameters
for distance dependence and $D$ is an adjustable parameter of the hydrogen bond energy.
For WHB  $D\approx 4 \div 6$ kcal/mol
at $\varepsilon =3.5$ with $D=5$ kcal/mol being a conventional value
\cite{Fin02} (in \cite{Pop97} $D$ is assumed to be a decreasing
value at increasing $\varepsilon$ with $D=0.5$ kcal/mol at
$\varepsilon =10$). Finally $r(x_i)$ is the current length of the
hydrogen bond.

We denote $a^{(j)}$ the distance from the axis $\sigma$ (see Fig.2) to the carbonyl oxygen atom ($j=1$ so that $a^{(1)}\approx 1.5$ \AA\ (see Appendix))
or that to the hydrogen atom ($j=2$ so that $a^{(2)}\approx 2$ \AA\ (see Appendix)). Then the current length
is related to the angle $x_i$ by straightforward trigonometry
\[
\left(r^{(j)}\left(x_i^{(j)}\right)\right)^2=\left(l^{(j)}\right)^2+4\left(a^{(j)}\right)^2
\left(\sin(x_i^{(j)}/4)\right)^2\left[a^{(j)}+l^{(j)}\cos \theta^{(j)}\right]
\]
This expression takes into account that the range for the angular displacement
$x_i$ (defined in (\ref{eq1})) is 4$\pi$ because it is twice of the
torsional angles $\varphi_i$ and $\psi_{i-1}$ (see (\ref{eq1})).
However we also want to take into account that the current length decreases with the increase of the
hydrogen bond energy. We attain it by introducing a linear dependence factor $\left (1-q D \right )$ in the expression for the current length where $q$ is an empirical parameter
\[
r^{(j)}\left(x_i^{(j)}\right)=\left (1-q D^{(j)}\right )\times 
\]
\[
\sqrt{\left(l^{(j)}\right)^2+4\left(a^{(j)}\right)^2
\left(\sin\left(x_i^{(j)}/4\right)\right)^2\left[a^{(j)}+l^{(j)}\cos \theta^{(j)}\right]}
\]
Thus for the ${U}^{(j)}_{hb}(x_i^{(j)})$ ($j$=1,2) we obtain
\[
U^{(j)}_{hb}(x_i^{(j)})=-D^{(j)}+D^{(j)}\Biggl \lbrace 1-\label{eq16} exp\Biggl [-n\Biggl (-r_0^{(j)}+
\left (1-q D^{(j)} \right )\times
\]
\begin{equation}
\label{eq16} \sqrt {\left(l^{(j)}\right)^2+4\left(a^{(j)}\right)^2
\left(\sin\left(x_i^{(j)}/4\right)\right)^2\left[a^{(j)}+l^{(j)}\cos \theta^{(j)}\right]}\Biggr )\Biggr]\Biggr \rbrace^2
\end{equation}
Here $n=3$ \AA$^{-1}$ and $r_0=1.8$ \AA\ are old adjustable parameters while $q$ and $l^{(j)}$ are
two new ones. The latter can be evaluated from the following data obtained
in \cite{Sma97}. A 0.5 \AA\  lengthening of the low-barrier hydrogen bond results in a weakening of that bond by over 6 kcal/mol.
A 1.0 \AA\  lengthening of the hydrogen bond results in an approximately 12 kcal/mol decrease in the calculated strength of the corresponding hydrogen bond. Taking the energy of the low-barrier hydrogen bond to be $D=15$ kcal/mol we obtain from these requirements and (\ref{eq16}) that $q\approx 0.01$  mol/kcal and $l^{(j)}\approx 2.0$ \AA. The angles $\theta^{(1)}$ and $\theta^{(2)}$ can be evaluated from the known data on the bond lengths and angles in the peptide group \cite{Can80} as $\theta^{(1)}\approx\theta^{(2)}\approx 30^{\circ}$.

\section{Effective on-site potential and coupling interaction potential}
The adjustable parameters are chosen to satisfy the following set
of requirements: 1. The effective on-site potential $V_{eff}(x_{i})$ (see \ref{eq3}) must
have a minimum at equilibrium ($x_i=0$) because the antiparallel $\beta$-sheet is
a steady stable structure. 2. For the anti-parallel $\beta $-sheet the hydrogen bond
are known (see, e.g., \cite{Bra99} or \cite{Can80}) to be practically parallel to the corresponding covalent bonds $C=O$ and $N-H$ (see Fig.2). 3. For the anti-parallel $\beta $-sheet the
spectroscopic frequency
\begin{equation}
\label{eq17}
(1/\lambda)_{sp}=\frac{\sqrt{V''_{eff}(x_i=0)/I}}{2\pi c}
\end{equation}
are $\sim  100\ cm^{-1}$. Here $\lambda$ is a wavelength, $c$ is
the light speed and the dash denotes a derivative in $x_i$.

We take for the anti-parallel $\beta $-sheet $\varphi _{i}^{0}=-139^{\circ}$;
$\psi _{i}^{0} =135^{\circ}$ \cite{Can80}. One hydrogen bond for the peptide group the anti-parallel $\beta $-sheet (namely that with $j=1$) always remains to be WHB ($D^{(1)}=5$ kcal/mol).
For the second hydrogen bond we vary the value $D^{(2)}$ for the energy within the range $2 \div 30$ kcal/mol.

In what follows we use dimensionless variables. We define the
frequency
\begin{equation}
\label{eq18} \omega=\sqrt{\frac{V''_{eff}(x_i=0)}{I}}
\end{equation}
and measure time in the units of $\omega^{-1}$ so that the
dimensionless time is
\begin{equation}
\label{eq19} \tau=t\omega
\end{equation}
Also we measure energy in the units of $I\omega^2$ and denote the
dimensionless coupling constant $\rho $
\begin{equation}
\label{eq20} \rho = \frac{K}{I\omega ^2}
\end{equation}

\section{Equation of motion and correlation function for crankshaft motion}
In this Sec. we denote for simplicity $x _{i}(t)\equiv \phi (t)$ and rewrite (\ref{eq1}) in the form
\begin{equation}
\label{eq21} \phi/2\equiv \Delta \varphi_i = \Delta \psi_{i-1}
\end{equation}
It is natural to describe the crankshaft motion by a Langevin equation.
Such equations are frequently used in protein dynamics
\cite{Ros79}, \cite{Dos06}, \cite{Rub04}, \cite{Sit96}, \cite{Sit00}, \cite{Sit02}.
The equation of the crankshaft motion for the rigid plane of the peptide
 group in the above conditions is
\begin{equation}
\label{eq22} I\frac{d^2 \phi(t)}{dt^2}+\gamma \frac{d \phi(t)}{dt}+
I\omega_0^2\phi(t)=\xi(t)
\end{equation}
where $\gamma$ is the friction coefficient and $\xi(t)$ is the random torque
with zero mean $<\xi(t)=0>$ and correlation function
\begin{equation}
\label{eq23} <\xi(0)\xi(t)>=2k_BT\gamma \delta(t)
\end{equation}
 $k_B$ is the Boltzman constant, $T$ is the temperature.

We consider the hydrodynamic friction. This "macroscopic"
notion is known to work surprisingly well at the molecular level
(see \cite{Met77} for thorough discussion).
We model the peptide group by an oblate
ellipsoid with half-axes $a$, $b$ and $c$ (where $a \sim c $
 and $a, c >> b$
 that reflects the flat character of the peptide group) Fig.8.
 For the rotation of the ellipsoid around $x$-axis
 (that is in our case actually
 the axis $\sigma$ for the peptide group introduced above)
the friction coefficient is given by a formula \cite{Klu97}
\begin{equation}
\label{eq24} (\gamma)_x=8\pi \eta \ abc\left[\frac{12 r}{a}+
\frac{(bc)^{1/4}}{a^{1/2}\left(1+
\frac{4r}{a^{1/2}(bc)^{1/4}}\right)^3}\right]^{-1}
\end{equation}
where $r=2(abc)^{1/3}$ and $\eta$ is the viscosity of the
 ellipsoid environment. The required behavior is obtained if we have
 the condition of the underdamped motion
\begin{equation}
\label{eq25} \frac{\gamma}{I\omega_0} << 1
\end{equation}
At such requirement we obtain (see e.g. \cite{Chir80}, \cite{Rub04})
\begin{equation}
\label{eq26} \alpha(t)\equiv <\phi(0)\phi(t)>\approx
 \frac{k_BT}{I\omega_0^2}exp\left(-\frac{\gamma \mid t \mid}{2I}\right)
cos\left(\omega_0t\right)
\end{equation}
We denote
\begin{equation}
\label{eq27} \mu=\frac{\gamma}{\sqrt 2 I\omega_0}
\end{equation}
Then the Fourier spectrum for the correlation function $\alpha(t)$ is
\begin{equation}
\label{eq28} \tilde \alpha(\omega)=\frac{k_BT}{I\omega_0^2}
\frac{\mu}{\pi}\frac{1+\mu^2+\left(\omega/\omega_0\right)^2}
{\mu^4+2\mu^2\left[1+\left(\omega/\omega_0\right)^2\right]+
\left[1-\left(\omega/\omega_0\right)^2\right]^2}
\end{equation}
At our requirement (\ref{eq25}) it has a sharp resonance peak
at the frequency $\omega_0$.
For the mean squared amplitude ($msa$) we have
\begin{equation}
\label{eq29} msa=\frac{k_BT}{I\omega_0^2}
\end{equation}
The latter means that the amplitude of the crankshaft motion at room
temperature and, e.g., typical value $\omega_0 \approx  10^{13} s^{-1}$ is
\begin{equation}
\label{eq30} \phi_{max}=\sqrt{\frac{k_BT}{I\omega_0^2}}
\approx 0.1\approx 6^{\circ}
\end{equation}
that is $\Delta \varphi_i = \Delta \psi_{i-1}\approx 3^{\circ}$.
This value sheds light on the origin of essentially vibrational character of
the peptide group motion manifested itself in (\ref{eq26}) and (\ref{eq28}).
 At such angles of
rotational deviation of the peptide group from its mean averaged position
the linear displacements of the atoms are
$\sim c \cdot \phi_{max}\approx 0.1\ \AA$ that is much less than
both the size
of a solvent molecule and interatomic distances to
neighbor fragments of protein structure. That is why the environment
exerts rather weak friction for such type of the peptide group motion that is
reflected in the requirement (\ref{eq25}). Thus we conclude that the
thermally equilibrium crankshaft motion of the peptide group is of
essentially vibrational character even in such
condensed medium as protein interior. Our model yields for the correlation function
of the crankshaft motion the exponentially decaying oscillations (\ref{eq26}).
This type of behavior is in qualitative agreement with the
results of molecular dynamics simulations \cite{Fit07}.

\section{Results and discussion}
The effective on-site potential for the peptide group in the anti-parallel $\beta $-sheet for the case of one weak hydrogen bond ($D^{(1)}=5$ kcal/mol) and one strong hydrogen bond ($D^{(2)}=30$ kcal/mol) is presented in Fig.4. In Fig. 5 the dependence of the spectroscopic frequency of crankshaft motion $(1/\lambda)_{sp}$ (see (\ref{eq17})) for the peptide group in the antiparallel $\beta$-sheet on the energy of one of its hydrogen bonds (namely $D^{(2)}$) at different values
of the torsional potentials for the rotation of the angles $\varphi_{i}$ and $\psi _{i-1}$ (see (\ref{eq14})) is depicted. This Fig. shows that at transformation WHB $\rightarrow$ SHB  of WHB with energy 5 kcal/mol into SHB with energy say $15\div20$ kcal/mol the frequency of crankshaft motion becomes roughly twice of what it was ($(1/\lambda)_{sp}\approx 40\ cm^{-1} \rightarrow\ (1/\lambda)_{sp}\approx 80\ cm^{-1}$). In Fig. 6 the dependence of the dimensionless coupling constant $\rho $  (see (\ref{eq20})) for the peptide group in the antiparallel $\beta$-sheet on the energy of one of its hydrogen bonds (namely $D^{(2)}$) at different values of the torsional potentials for the rotation of the angles $\varphi_{i}$ and $\psi _{i-1}$ (see (\ref{eq14})) is depicted. The dimensionless coupling constant decreases in absolute value with the increase of $D^{(2)}$ because it is inversely proportional to the $\omega^2$ (see (\ref{eq20})) while the latter increases with the increase of $D^{(2)}$ (see Fig. 6). The dimensional value of the coupling constant $K$ remains unchanged with variation of $D^{(2)}$ because does not depend on the hydrogen bond potential (see (\ref{eq12})).
In Fig.7 the dependence of the coupling constant $\rho $  (see (\ref{eq20})) for the peptide group in the antiparallel $\beta$-sheet on the dielectric constant $\epsilon$ at different values of the energy of one of its hydrogen bonds (namely $D^{(2)}$) is depicted.
This Fig. shows that the dependence of the coupling constant $\rho $  (see (\ref{eq20})) for the peptide group in the antiparallel $\beta$-sheet on the dielectric constant $\epsilon$ is rather strong.
On the other hand the dependence of the frequency of crankshaft motion for the peptide group in the antiparallel $\beta$-sheet on the dielectric constant $\epsilon$ is negligibly small.
This can be explained as follows. The frequency of crankshaft motion is determined by the
the effective on-site potential (see \ref{eq3} and \ref{eq13}) that does not include electrostatic interactions while the coupling constant is determined by the coupling interaction potential (see \ref{eq12} and \ref{eq20}) that strongly depends on electrostatic interactions and thus very sensitive to the value of $\epsilon$.

Our results of the investigation of the effective on-site
potential and the coupling interaction potential for the anti-parallel $\beta $-sheet testify the following: 1. For the anti-parallel $\beta $-sheet the
coupling interaction is surprisingly weak $\mid \rho \mid << 1$ (though the same is true for other types of protein secondary structure \cite{Sit07}). 2. The effective on-site potential
is hard (goes more steep than a harmonic one). 3. The coupling
interaction is repulsive $\rho < 0$ at reasonable values of the dielectric constant $\epsilon < 15$ (the latter means that a non-zero value of a peptide group
displacement tends to increase the values of the neighboring
peptide groups displacements with the opposite sign). The weakness of the coupling interaction $\mid \rho \mid << 1$ is known to be indispensable for the possibility of creation and existence of a specific excitation (the so-called discrete breather) in a chain of non-linear oscillators. Thus we conclude that
in the anti-parallel $\beta $-sheet a discrete breather can be excited.

The fact that dimensionless values of the coupling constant $\rho $ are very small, i.e., $\mid \rho \mid  << 1$ (see Fig.6 and Fig.7)
means that there is little interdependence in the rocking (i.e., Crankshaft motion $\Delta \varphi_i = \Delta \psi_{i-1}$) of the planes of the adjacent peptide groups in the antiparallel $\beta$-sheet. This result is in agreement with the conclusion of the authors of \cite{Fit07} that "only a slight correlation is found between the motions of the two backbone dihedral angles of the same residue" (i.e., $\Delta \varphi_i$ practically does not correlate with $\Delta \psi_{i}$).

The most intriguing question is how the the transformation of
WHB into SHB in the antiparallel $\beta$-sheet of the nonspecific binding site of serine proteases can be related to the mechanism of enzyme action? The hypothesis that such relationship can take place was suggested in \cite{Fod06} (see Introduction). In our opinion it is very interesting and requires discussion and further development. The indispensable first step is the storage of the energy released at such transformation. The second step is the transmission of this energy to the scissile bond in the substrate molecule. There are several possible mechanisms of such storage in the antiparallel $\beta$-sheet. One of them makes use of the phenomenon of the so-called discrete breather. This possibility was discussed in \cite{Sit07}, \cite{Sit06}. The energy of the discrete breather can be transmitted to the scissile bond in the substrate molecule by the electric field fluctuations \cite{Sit06}. Another interesting possibility is the excitation at the WHB $\rightarrow$ SHB transformation of the self-trapped Amide-I vibration in the $C=O$ stretching mode of the nearest to the active center peptide group in the antiparallel $\beta$-sheet of the nonspecific binding site of serine proteases. Indeed the Amide-I quantum energy is known to be 0.205 eV (see, e.g., \cite{Can80}), i.e., $\ \approx 4.7$ kcal/mol. Thus the transformation of a WHB $N-H \cdot\cdot\cdot\cdot\cdot O=C$ with energy 5 kcal/mol into a low-barrier hydrogen bond $N-H \bullet\bullet\bullet O=C$ with energy, e.g, 15 kcal/mol releases enough energy to excite Amide-I of the $C=O$ bond. Of course one should keep in mind that there can be complications due to the fact that the bond between $C$ and $O$ atoms in a peptide group is only partially double one (see, e.g.,\cite{Can80} or \cite{Bra99}). However these complications do not seem to be crucial. If the $C=O$ vibration is excited
then the energy can be transmitted to the scissile bond $C_{carbonyl}-N$ in the peptide group of the substrate molecule by the nonradiative resonant interaction of the dipole moment $\bar d$ of the stretching mode with that of the $C_{carbonyl}=O$ bond adjacent to $C_{carbonyl}-N$. At the oversimplifying approximation that both dipoles have equal tilt relative the axis $z$ connecting them the exchange energy (vibrational coupling) of the dipole-dipole interaction $J$ can be evaluated as \cite{Dav79}, \cite{Dav84}
\[
J=\frac{\bar d^2}{R^3}\left(3\cos^2\theta-1\right)
\]
Here $R$ is the distance between the $C=O$ bonds and $\theta$ is the angle between the dipole moment $\bar d$ and the axis $z$. The value of $J$ for, e.g., peptide groups in $\alpha-$helix was estimated from experimental data of infrared spectra and it was found that $J=1.55\cdot10^{-22}$ Joul \cite{Nev76}, i.e., $J\approx 2.22\cdot10^{-2}$ kcal/mol. Thus the Hamiltonian for two interacting $C=O$ bonds can be written as
\[
H_{ex}=\hbar\omega_0\sum_{n=1}^2 a_n^+a_n-J\sum_{n=1}^2 \left(a_n^+a_{n+1}+a_na^+_{n+1}\right)
\]
where $\omega_0$ is the frequency of Amide-I ($\sim 1600\ cm^{-1}$) and $a_n^+(a_n)$ is the creation (annihilation) operator for an Amide-I exciton in the site $n$.
The storage of vibrational energy in the $C=O$ stretching mode is widely used within the framework of the Davydov model for the bio-energy transport in a polypeptide chain \cite{Dav79}, \cite{Dav84}. In this model the $C=O$ vibration is excited by the energy released in the adenosine triphosphate (ATP) hydrolysis that yields about 0.43 eV. Hamiltonian of the above type is still a matter intensive investigations within the context of energy transfer in proteins (see, e.g., \cite{Kob11} and refs. therein).
Regretfully we failed to find experimental or theoretical investigations of the key point for our hypothetical scenario namely that of Amide-I excitation at transformation of the ordinary hydrogen bond into low-barrier ones. That is why we restrict ourselves from further discussing this speculation and leave the development of posed above possibility for future work.

The conclusions of the paper are summarized as follows. The dynamics of a peptide chain in the anti-parallel $\beta $-sheet is considered within a realistic model
with stringent microscopically derived coupling interaction
potential and effective on-site potential. The coupling
interaction is found to be surprisingly weak
and repulsive in character. The effective on-site potential is found to be a hard one. At transformation of ordinary hydrogen bond into low-barrier one the frequency of crankshaft motion of the corresponding peptide group in the anti-parallel $\beta $-sheet is roughly doubled. \\

Acknowledgements.  The author is grateful to Prof. Yu.F. Zuev and Dr. B.Z. Idiyatullin for
helpful discussions. The work was supported by the grant from RFBR and
the programme "Molecular and Cellular Biology" of RAS.

\section{Appendix A}
Here we give a brief estimate of the moment of inertia of the peptide group
for rotation around the axis $\sigma$ that goes through the
center of masses of the peptide group parallel to the bonds
$C_{\alpha}^{i-1}-C^i$ and $N^i-C_{\alpha}^{i}$ (see Fig.2).
The peptide group is a rigid plane structure so that all atoms $O$, $C$, $N$ and $H$
lie in a plane. That is why the moment of inertia is simply the sum of their
masses multiplied by the square of their distances to the axis $\sigma$.
The masses in atomic units ($1\ a.u.m.=1.7 \cdot 10^{-24} g$) are $m_O=16$,
$m_C=12$, $m_N=14$ and $m_H=1$. That is why the center of masses is shifted
a little to the atoms $C$ and $O$ relative the middle of the bond $C-N$. The
lengths of the bonds are $l_{OC}=1.24$ \AA, $l_{CN}=1.32$ \AA\ and
 $l_{NH}=1.0$ \AA. That is why the distances to the axis $\sigma$ are
approximately  $r_{O}\approx 1.5$ \AA, $r_{C}\approx 0.3$ \AA\ and
 $r_{N}\approx 0.7$ \AA\ and $r_{H}\approx 2$ \AA. Substituting these values into the
formula $I=\sum m_i r_i^2$ we obtain  $I \approx 7.6 \cdot 10^{-39}
g \cdot cm^{2}$. Thus our rough estimate corroborate the  precise value
 $I \approx 7.34 \cdot 10^{-39}
g \cdot cm^{2}$ from \cite{Flu94}.

\newpage

\newpage
\begin{figure}
\begin{center}
\includegraphics* [width=\textwidth]{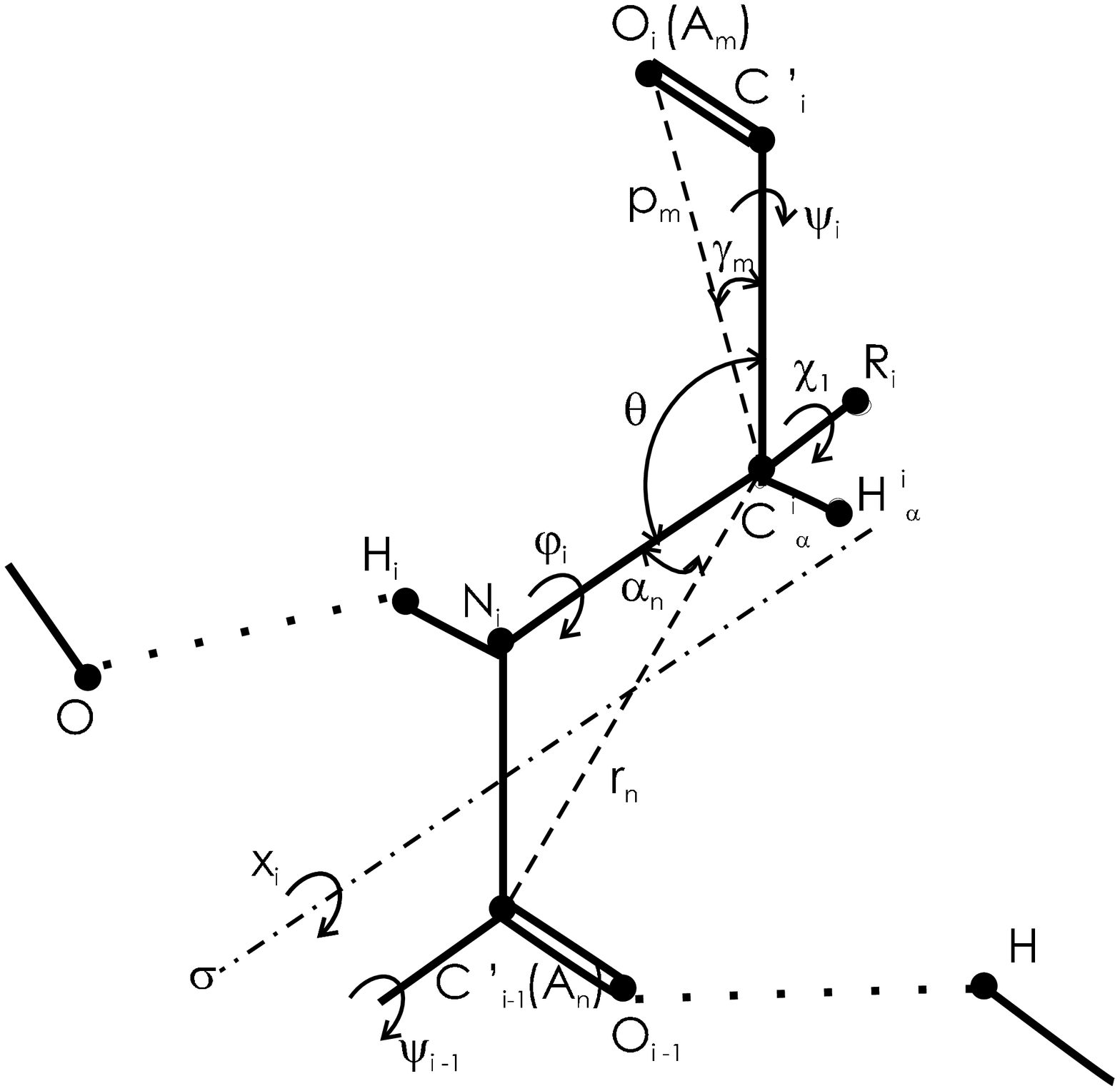}
\end{center}
\caption{A schematic representation of the peptide chain with all
designations necessary for the text. The covalent bonds are shown
as bold lines. Dotted lines denote hydrogen bonds. Dashed lines
denote auxiliary axes.} \label{Fig.1}
\end{figure}

\begin{figure}
\begin{center}
\includegraphics* [width=\textwidth] {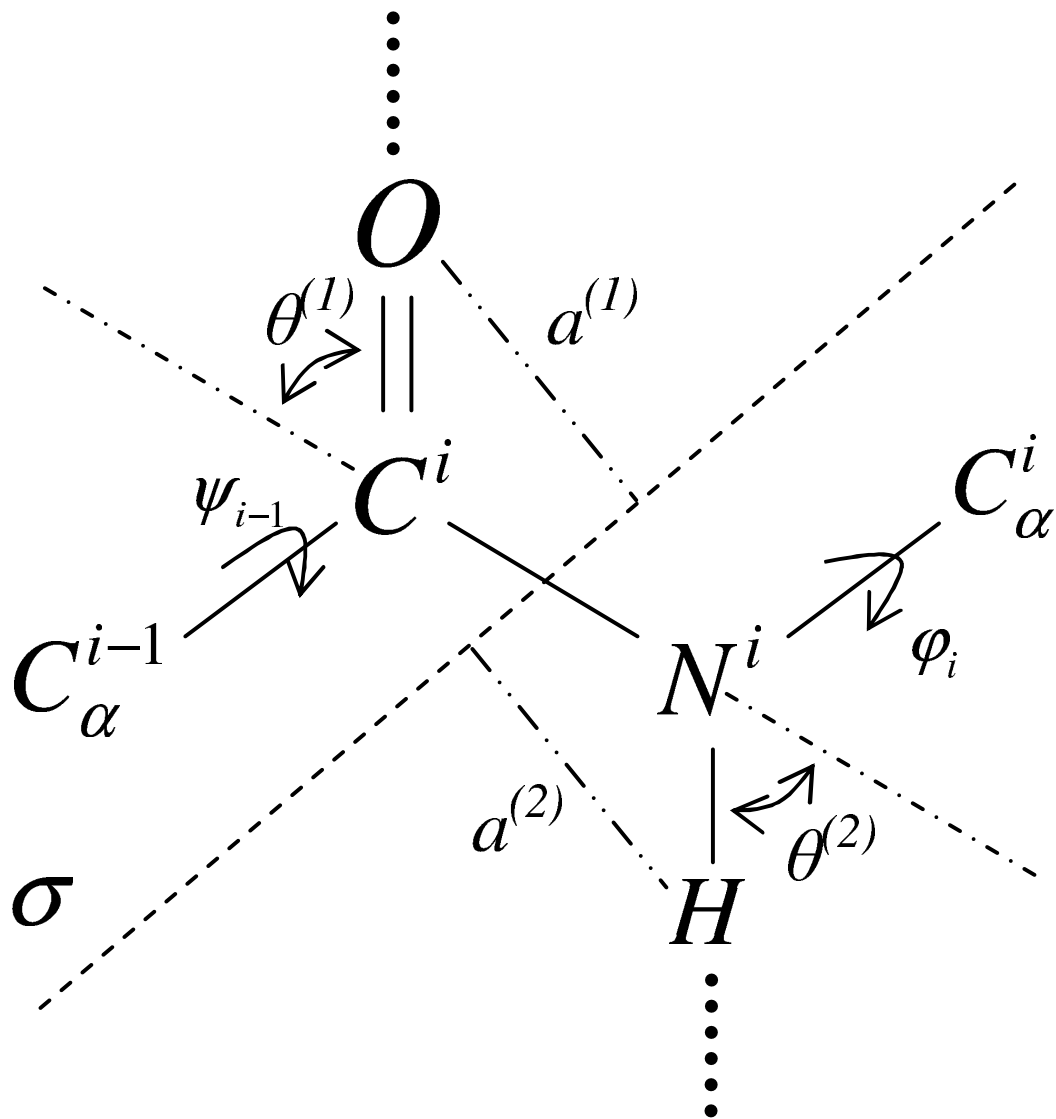}
\end{center}
\caption{Schematic picture of the peptide group in the antiparallel $\beta$-sheet .}
\label{fig:Fig.2}
\end{figure}

\clearpage
\begin{figure}
\begin{center}
\includegraphics* [width=\textwidth] {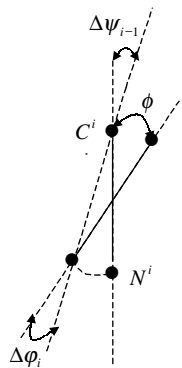}
\end{center}
\caption{A look on the peptide group from the axis of rotation $\sigma$
explaining the definition of the angle $\phi$ (defined in
(\ref{eq3})).}
\label{Fig.3}
\end{figure}

\clearpage
\begin{figure}
\begin{center}
\includegraphics* [width=\textwidth] {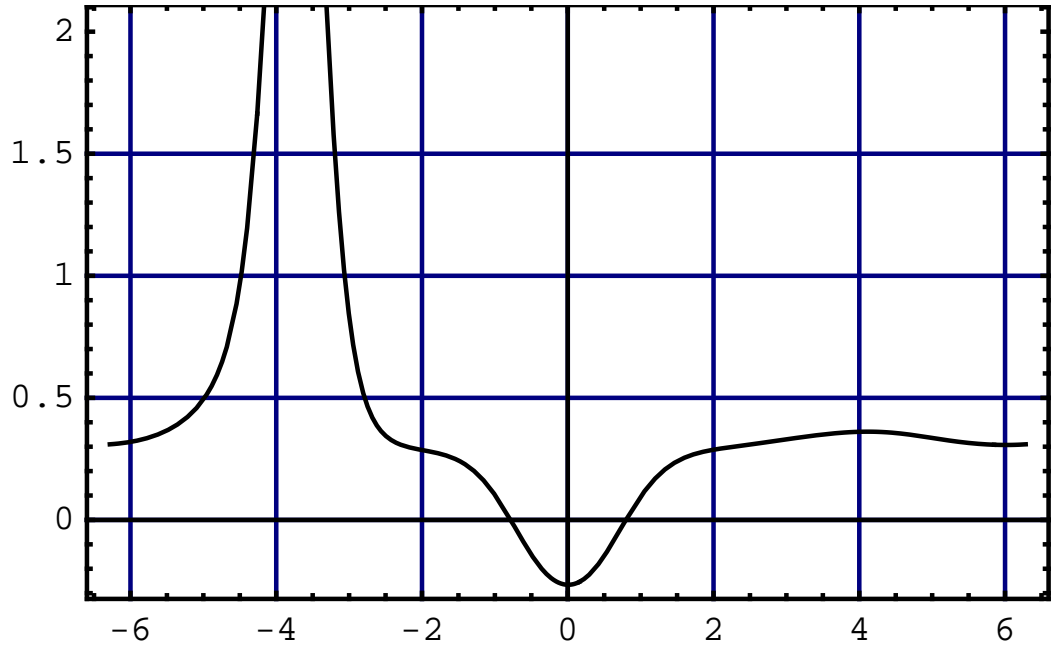}
\end{center}
\caption{The normalized effective on-site potential $V_{eff}(x)$
(see (\ref{eq3})) for the peptide group in the antiparallel $\beta$-sheet in case of one weak hydrogen bond ($D^{(1)}=5$ kcal/mol) and one strong hydrogen bond ($D^{(2)}=30$ kcal/mol). The value of the dielectric constant is $\varepsilon =3.5$. Here I is the moment of inertia of the peptide group relative the axis $\sigma$ and
$\omega$ is the frequency. The range for the angular displacement
$x$ (defined in (\ref{eq1})) is 4$\pi$ because it is twice of the
torsional angles $\varphi$ and $\psi$ (see (\ref{eq1})).} \label{Fig.4}
\end{figure}

\clearpage
\begin{figure}
\begin{center}
\includegraphics* [width=\textwidth] {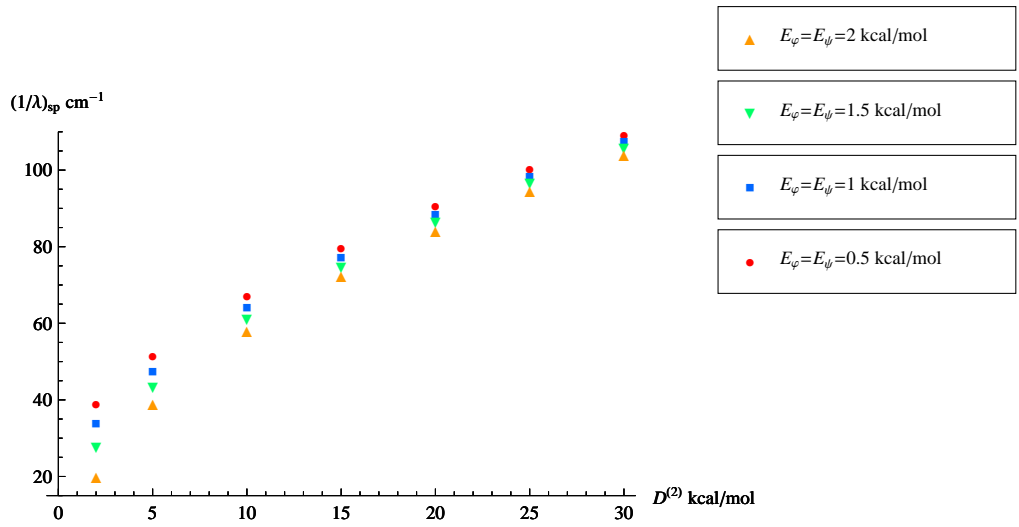}
\end{center}
\caption{The dependence of the spectroscopic frequency of crankshaft motion $(1/\lambda)_{sp}$
(see (\ref{eq17})) for the peptide group in the antiparallel $\beta$-sheet on the energy of one of its hydrogen bonds (namely $D^{(2)}$) at different values
of the torsional potentials for the rotation of the angles $\varphi
_{i}$ and $\psi _{i-1}$ (see (\ref{eq14})): $E_{\varphi}=$0.5 kcal/mol, $E_{\psi}=$0.5
kcal/mol; $E_{\varphi}=$1. kcal/mol, $E_{\psi}=$1.
kcal/mol; $E_{\varphi}=$1.5 kcal/mol, $E_{\psi}=$1.5
kcal/mol; $E_{\varphi}=$2. kcal/mol, $E_{\psi}=$2.
kcal/mol. The value of the dielectric constant is $\varepsilon =3.5$.} \label{Fig.5}
\end{figure}

\clearpage
\begin{figure}
\begin{center}
\includegraphics* [width=\textwidth] {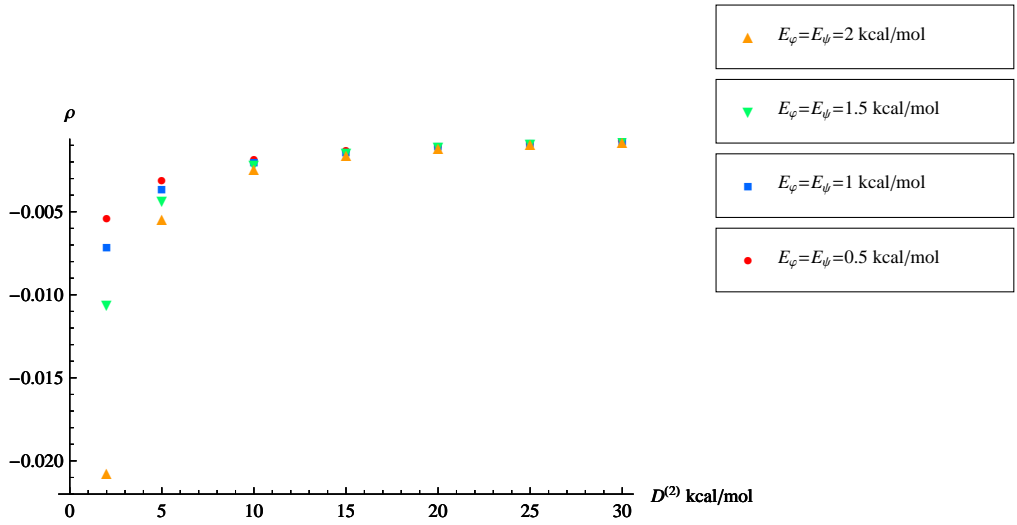}
\end{center}
\caption{The dependence of the coupling constant $\rho $  (see (\ref{eq20})) for the peptide group in the antiparallel $\beta$-sheet on the energy of one of its hydrogen bonds (namely $D^{(2)}$) at different values of the torsional potentials for the rotation of the angles $\varphi
_{i}$ and $\psi _{i-1}$ (see (\ref{eq14})): $E_{\varphi}=$0.5 kcal/mol, $E_{\psi}=$0.5
kcal/mol; $E_{\varphi}=$1. kcal/mol, $E_{\psi}=$1.
kcal/mol; $E_{\varphi}=$1.5 kcal/mol, $E_{\psi}=$1.5
kcal/mol; $E_{\varphi}=$2. kcal/mol, $E_{\psi}=$2.
kcal/mol. The value of the dielectric constant is $\varepsilon =3.5$.} \label{Fig.6}
\end{figure}

\clearpage
\begin{figure}
\begin{center}
\includegraphics* [width=\textwidth] {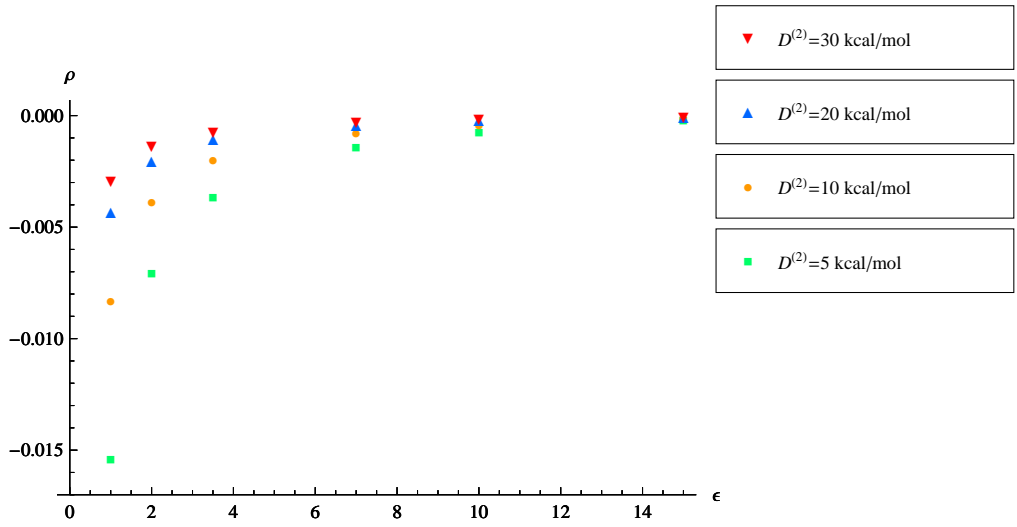}
\end{center}
\caption{The dependence of the coupling constant $\rho $  (see (\ref{eq20})) for the peptide group in the antiparallel $\beta$-sheet on the dielectric constant $\epsilon$ at different values
of the energy of one of its hydrogen bonds (namely $D^{(2)}$): $D^{(2)}=$5 kcal/mol; $D^{(2)}=$10 kcal/mol; $D^{(2)}=$20 kcal/mol; $D^{(2)}=$30 kcal/mol. The values of the torsional potentials for the rotation of the angles $\varphi
_{i}$ and $\psi _{i-1}$ (see (\ref{eq14})) are $E_{\varphi}=$1. kcal/mol, $E_{\psi}=$1.
kcal/mol. } \label{Fig.7}
\end{figure}

\clearpage
\begin{figure}
\begin{center}
\includegraphics* [width=\textwidth] {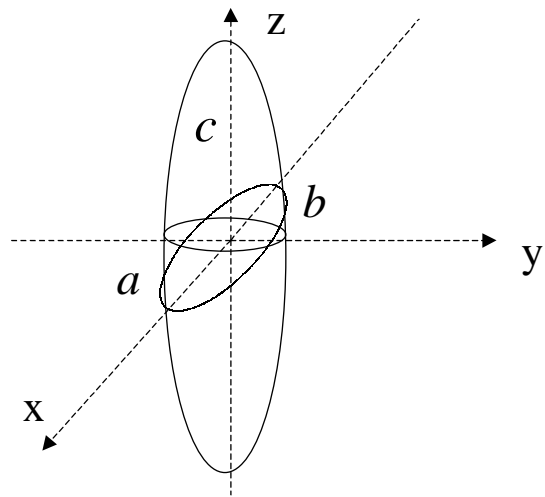}
\end{center}
\caption{Model of the peptide group by oblate ellipsoid.}
\label{Fig.8}
\end{figure}

\end{document}